\documentclass[a4paper,11pt]{article}

\usepackage{jcappub}
\usepackage[T1]{fontenc}
\usepackage[normalem]{ulem}
\usepackage{amsmath}
\usepackage{amssymb}
\usepackage{geometry}
\usepackage{booktabs}
\usepackage{array}
\usepackage{paralist}
\usepackage{verbatim}
\usepackage{graphicx}
\usepackage[utf8]{inputenc}
\usepackage{multirow}
\usepackage[dvipsnames*,svgnames]{xcolor}
\usepackage{wasysym}
\usepackage{subfigure}
\usepackage{enumitem}
\usepackage{lineno}
\usepackage{float}

\usepackage{lineno}  

\title{\textbf{Effects of Lorentz invariance violation on charged particles and photon production in astrophysical sources}}
\author{Matheus Duarte and}
\emailAdd{matheus\_duarte@usp.br}

\author{Vitor de Souza}
\emailAdd{vitor@ifsc.usp.br}

\affiliation{São Carlos Institute of Physics, University of São Paulo, Av. Trabalhador S\~ao-carlense 400, S\~ao Carlos, Brazil}

\abstract{We investigate the impact of Lorentz invariance violation (LIV) on radiation processes in astrophysical sources, focusing on synchrotron and inverse Compton interactions. We derive modified expressions for radiated power and photon energy under LIV assumptions and incorporate them into first-order Fermi acceleration models. Our analysis reveals energy thresholds beyond which LIV, within a kinematic framework, significantly alters particle dynamics and photon spectra, introducing non-physical divergences that highlight limitations in perturbative approaches. We model synchrotron self-Compton (SSC) emission in the presence of LIV and assess its consequences for photon fluxes from blazars, including Markarian 501 and the BL Lac population. LIV introduces distinct high-energy emission regions that deviate from standard expectations. Comparisons with observational data, particularly upper limits from the Pierre Auger Observatory, suggest that future multi-messenger observations, together with the full analysis of particle's trajectories, could constrain LIV parameters through the non-detection of such excesses.}

\begin{document}

\maketitle
\flushbottom

\section{Introduction}

Lorentz invariance (LI) is a fundamental principle in modern physics, underlying the theory of relativity~\cite{Einsteinrelativity} and the standard model of particle physics~\cite{Gaillard_1999,pdg}. It asserts that the laws of physics remain the same across all inertial frames of reference. However, potential violations of Lorentz invariance (LIV)~\cite{Mattingly_2005} have been theorized within the context of quantum gravity~\cite{Batista_2023,Addazi_2021,Ashtekar_2021} and other theories beyond the standard model~\cite{Greenberg_2002,Bernardini_2008}. Investigating these possible violations provides insight into fundamental aspects of the Universe, potentially revealing new physics at extremely high energies.

The propagation of astroparticles in the Universe~\cite{Carmona_2024,Amelino-Camelia:1997ieq,Jacob_2008,Schreck_2017,Zhang_2015} and in the development of extensive air showers~\cite{Saveliev_2023,Duenkel_2021} have imposed limits on LIV parameters~\cite{huerta,Satunin_2019,Albert_2020,Kosteleck__2015,Abreu_2022}. Nonetheless, a thorough investigation of the processes leading to the production of charged particles and photons under the hypothesis of LIV remains necessary~\cite{Duarte_2024}.

The mechanisms generating these charged particles and photons are fundamentally dependent on LI, which plays a critical role in determining the maximum energy, the emitted power, and the resulting energy spectrum~\cite{Ghisellini_2013}. Consequently, such sources provide a natural framework for probing potential modifications of LI. In particular, the detection of highly energetic astroparticles accompanied by photon counterparts offers a promising framework for probing potential deviations from LI within the context of multi-messenger astrophysics.

In this work, we investigate how LIV changes synchrotron and inverse Compton~\cite{Klein_1929} interactions, the two most important energy loss processes in astrophysical sources~\cite{Ridgers_2014,Sun_2023,Ruszkowski_2023,Longair_2011,Pfutzner_2023,Lenain_2008,Wen_2024}. We also incorporate the LIV models for synchrotron and inverse Compton interactions into the context of first-order Fermi acceleration and photon emission in blazars. We investigate how the production of high-energetic charged particles and photons is affected by the LIV assumption.

In Section~\ref{sec:liv}, we present the LIV framework. In Subsection~\ref{subsec:sync:theory}, we derive the expressions for the LIV-modified synchrotron radiated power and the corresponding synchrotron photon energy. The application of LIV to the inverse Compton scattering process is addressed in Subsection~\ref{subsec:ic:theory}. A combined treatment of synchrotron and inverse Compton emission within a synchrotron self-Compton (SSC) scenario, incorporating LIV effects, is developed in Section~\ref{sec:firstorder}, where we analyze a case study of first-order Fermi acceleration. In this context, we independently evaluated the impact of LIV on both energy losses and energy spectrum of particles, as well as on the resulting photon emission. Appendix~\ref{sec:blazar} presents an illustrative comparison between our theoretical predictions and observational data, including the identified blazar population and upper limits on photon flux from the Pierre Auger Observatory. Finally, our conclusions and implications are summarized in Section~\ref{sec:conclusion}.

\section{LIV framework}
\label{sec:liv}

Lorentz Invariance Violation (LIV), as predicted by various theoretical models~\cite{Tasson_2014,kolstelecky_2008,Jacobson_2006}, leads to a modification of Einstein's energy-momentum dispersion relation. This deviation can be expressed by additional terms in the standard dispersion relation:
\begin{equation}
    E^2 = m^2 + p^2 + \sum_n \delta_n p^{n+2} \rm{,}
    \label{eq:liv}
\end{equation}
where we have set $c=1$. The momentum coefficients $\delta_n$ represent the parameters responsible for the strength of LIV, while $n$ denotes the order of violation. Considering that we focus on highly energetic particles, we assume $p \gg m$, which leads to
\begin{equation}
    p = \frac{E}{\sqrt{1+\delta_nE^n}} \rm{.}
    \label{eq:liv:p}
\end{equation}
No preferred reference frame is assumed, so the modified dispersion relation applies in every frame \cite{Bluhm:2004ep,Kostelecky:2003fs, Graesser:2005bg}.

Lorentz Invariance Violation can also lead to modifications in the Lorentz factor, which are naturally obtained by considering the group velocity. A related discussion in the context of deformed special relativity can be found in~\cite{Morais:2023amp}. To obtain this modification, one may start from the definition of the Lorentz factor, $\gamma^2 = (1-v^2)^{-1}$, which, when combined with the group velocity, $v = \partial E /\partial p$ \cite{Amelino-Camelia:1997ieq, Mattingly_2005}, leads to
\begin{equation}
\gamma^2 = \frac{1}{1-(\frac{\partial E}{\partial p})^2} = \frac{E^2}{E^2 - p^2 - (n+2)\delta_n p^{n+2} - \left(\tfrac{n+2}{2}\right)^2 (\delta_n)^2 p^{2n+2}} \ \textrm{.}
\end{equation}
Recalling that $E^2 - p^2 = m^2 + \delta_n p^{n+2}$, this expression can be written as 
\begin{equation} 
\gamma^2 = \frac{E^2}{m^2 - (n+1)\delta_n p^{n+2} - (\frac{n+2}{2})^2 (\delta_n)^2 p^{2n+2}} \, \textrm{.} 
\end{equation}
It should be noted that this perturbative treatment may yield unphysical results for the modified factor (such as negative values or divergences). Therefore, we restrict our analysis to the parameter domain of $\delta_n$ in which the Lorentz factor remains physically meaningful. Moreover, the correction term $\delta_n p^{n+2}$ dominates over the quadratic contribution $(\delta_n p^{n+1})^2$ up to scales of $p^n \sim (\delta_n)^{-1}$. However, in this extreme regime the Lorentz factor would already become negative, thus falling outside the physically acceptable domain. Consequently, we can safely neglect the quadratic term within the energy and $\delta_n$ ranges of interest.

Under these considerations, and in the high-energy limit, we obtain the final expression for the Lorentz factor modified by LIV:
\begin{equation} 
\gamma^2_{\textrm{LIV}} = \frac{E^2}{{m^2 -\left(n+1 \right)\delta_nE^{n+2}}} \textrm{.} 
\label{eq:liv:factor} 
\end{equation} 
From equation~\ref{eq:liv:factor}, the allowed domain for the values of $\delta_n$ becomes evident. For each fixed pair of mass and energy, the LIV parameter is restricted to the region $\delta_n < \frac{m^2}{(n+1)E^{n+2}}$. For values outside this domain, the Lorentz factor diverges or becomes negative, revealing a fundamental limitation of the current model and indicating the need for a non-perturbative treatment.

Different particles may exhibit different breaking parameters~\cite{Mattingly_2005}. Therefore, we will denote the breaking parameters for electrons as $\delta_n^{(e)}$, for protons as $\delta_n^{(p)}$ and for photons as $\delta_n^{(\gamma)}$. When $\delta_n$ is used without specifying the type of particle, a charged particle is being considered.
\section{Radiation processes in particle acceleration under LIV assumption}

Accelerated charged particles emit electromagnetic radiation as described by Larmor's formula~\cite{Ghisellini_2013,Jackson_1998,Griffiths_2017}. The emitted power is
\begin{equation}
    P = \frac{2}{3}q^2\gamma^6\left[ |\dot{\Vec{v}} |^2 - |\Vec{v}\times \dot{\Vec{v}}| \right]\rm{,}
    \label{eq:larmor}
\end{equation}
where $q$ is the charge of the particle, $v$ and $\gamma$ are, respectively, its velocity and its Lorentz factor in the laboratory frame of reference. In the next subsections, we solve synchrotron and inverse Compton radiation considering LIV.

\subsection{Synchrotron radiation under LIV assumption}
\label{subsec:sync:theory}

A charged particle moving with velocity $\vec{v}$ through a magnetic field ($\vec{B}$) is subject to the Lorentz force, $\vec{F} = q \vec{v} \times \vec{B}$. This acceleration results in the emission of electromagnetic radiation, a process known as synchrotron radiation. Given that the force is always perpendicular to the velocity $\vec{F} = q \vec{v} \times \vec{B} = \gamma m \dot{\vec{v}}$, the power emitted in synchrotron radiation is 
\begin{equation}
    P = \frac{2}{3} \frac{q^4}{m^2} \gamma^2 |\vec{v} \times \vec{B} |^2 \, .
    \label{eq:larmor2}
\end{equation}
The energy of the emitted photons is highly concentrated at the peak of the spectral power distribution~\cite{Pacho_1970}. For simplicity, we assume the charged particle emits all photons with the same energy at the peak frequency 
\begin{equation}
E_{\gamma} = 0.29 \cdot \frac{3}{2} \gamma^2 \frac{qB}{m} \, .
\label{eq:peak}
\end{equation}
The previous equations can be used to analyze synchrotron emission when LIV is considered. Here we adopt a kinematic approach, in which LIV effects are included only through the modification of the Lorentz factor and the corresponding photon energy. This method does not alter the underlying quantum electrodynamics (QED) interaction vertices, and is intended as a first-order approximation to explore potential LIV effects on synchrotron emission arising exclusively from the modified dispersion relation. Therefore, using equations~\ref{eq:larmor2} and \ref{eq:peak}, along with \ref{eq:liv:factor}: 
\begin{equation}
    P^{\text{LIV}} = \frac{2}{3} \frac{q^4}{m^4} |\vec{v} \times \vec{B} |^2 \left[\frac{E^2}{1 -\left(n+1 \right)\frac{\delta_n}{m^2}E^{n+2}} \right]
    \label{eq:pot:sync}
\end{equation}
and
\begin{equation}
    E_{\gamma}^{\text{LIV}} =0.29 \cdot \frac{3}{2}\cdot \frac{qB}{m^3}\left[\frac{E^2}{1 -\left(n+1 \right)\frac{\delta_n}{m^2}E^{n+2}}\right] \rm{.}
    \label{eq:photon:sync}
\end{equation}
LIV modifies the Lorentz factor, which in turn affects the synchrotron power and characteristic photon energy. In the present kinematic framework, these equations capture the leading-order effects of LIV on the emission spectrum. For particles with energy $E^{n+2} > \frac{m^2}{(n+1)\delta_n}$, the perturbative approach breaks down, signaling that a full dynamical treatment, possibly including modified interaction vertices, would be required to fully describe such extreme events. However, as we are restricting the analysis to the physically allowed dominion of $\delta_n$, the present approach provides a reasonable first-order estimate of kinematic LIV effects on the synchrotron emission.

Similar qualitative conclusions regarding the leading effect of LIV on synchrotron emission were obtained in full LIV-QED frameworks~\cite{Altschul:2005za,Montemayor:2005ka}.

\subsection{Inverse Compton under LIV assumption}
\label{subsec:ic:theory}

Inverse Compton scattering is the interaction in which a low-energy photon gains energy by scattering off a high-energy charged particle. The energy of the scattered photon can be calculated imposing the conservation of the Lorentz-invariant 4-momentum 
\begin{equation}
p_p^{\mu} + p_{\gamma}^{\mu} = p_p^{'\mu} + p_{\gamma}^{'\mu} \textrm{,}
\end{equation}
where $p$ denotes the charged particle, $\gamma$ denotes photon and primed quantities refer to final states. 

Within our kinematic framework, the scattering cross section of relativistic particles is given by the Klein-Nishina formula~\cite{Klein_1929}
\begin{equation}
\sigma_{\textrm{KN}} = \frac{r_e^2}{2}\int \left( \frac{E'_\gamma}{E_\gamma}\right)^2\left( \frac{E'_\gamma}{E_\gamma} + \frac{E_\gamma}{E'_\gamma} - \sin^2{\theta} \right)d\Omega \textrm{,}
\label{eq:klein}
\end{equation}
and the average scattering angle can be calculated through
\begin{equation}
\left< \cos{\theta} \right> = \frac{\int \cos{\theta}\frac{d\sigma}{d\Omega} d\Omega}{\int \frac{d\sigma}{d\Omega} d\Omega} \textrm{,}
\label{eq:average}
\end{equation}
such that a full dynamic treatment could introduce additional corrections to the cross-section \cite{Rubtsov:2012kb, Carmona_2024} and possibly alter the scattering angle distribution.

As we restrict our analysis to a kinematic approach with no preferred reference frame, LIV can be introduced via the energy-momentum in Equation~\ref{eq:liv} in the charged particle frame of reference. With this modification, the spatial components ($\mu = 1,2,3$) become
\begin{multline}
    E^{'2}_p = m^2 + \left(\frac{E_\gamma^{'2}}{1+\delta_n^{(\gamma)}E_\gamma^{'n}} +  \frac{E_\gamma^{2}}{1+\delta_n^{(\gamma)}E_\gamma^n} - \frac{2E_\gamma E_\gamma^{'}\cos{\theta}}{\sqrt{\left(1+\delta_n^{(\gamma)}E_\gamma^{'n}\right) \left( 1+\delta_n^{(\gamma)}E_\gamma^n\right)}} \right) \times \\
    \times \left[1+\delta_n \left(\frac{E_\gamma^{'2}}{1+\delta_n^{(\gamma)}E_\gamma^{'n}} +  \frac{E_\gamma^{2}}{1+\delta_n^{(\gamma)}E_\gamma^n} - \frac{2E_\gamma E_\gamma^{'}\cos{\theta}}{\sqrt{\left(1+\delta_n^{(\gamma)}E_\gamma^{'n}\right) \left( 1+\delta_n^{(\gamma)}E_\gamma^n\right)}} \right)^{n/2}\right]^{1/2} \textrm{,}
    \label{eq:final:energy:collision}
\end{multline}
where $\theta$ is the scattering angle, while the time component ($\mu = 0$) yields
\begin{equation}
    m^2 + 2mE_\gamma + E^{2}_\gamma - E^{'2}_\gamma - E_p^{'2} - 2E'_\gamma E'_p = 0 \, \textrm{.}
    \label{eq:liv:ic}
\end{equation}

Reference~\cite{Abdalla_2018} showed similar results for the case $n=1$. However, here, we present a generalization for all values of $n$. For this interaction, LIV effects could, in principle, appear in both the photon and charged-particle sectors. In this work, we focus on the photon sector, where the modified dispersion relation directly affects the kinematics of the scattering process. Although several studies have derived strong bounds on LIV coefficients for charged particles, mainly from the non-observation of vacuum Cherenkov radiation~\cite{Saveliev:2024whq, Schreck_2017, Kaufhold:2005vj, Petrov:2025wey}, such analyses typically treat each sector independently and neglect combined propagation and cascade effects, a methodological simplification that we also adopt here to isolate the photon contribution.

\begin{figure}[ht]
    \centering
    \includegraphics[width=0.8\textwidth]{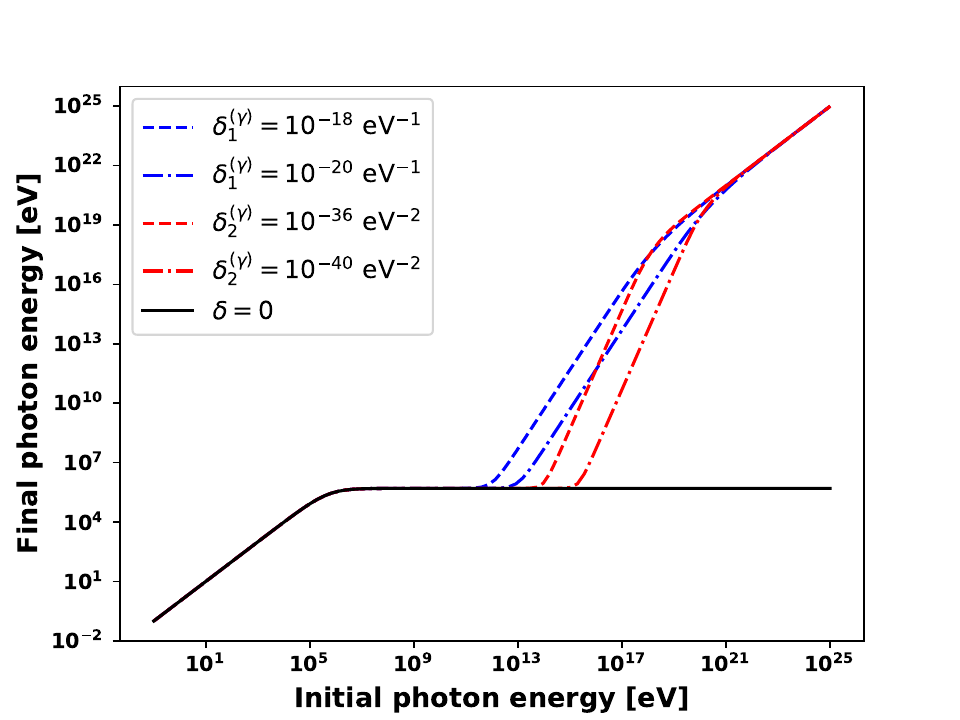}
    \caption{Final photon energy in the reference frame of the electron as a function of its initial energy. A scattering angle of $\pi/2$ was adopted. Different values of $\delta_n^{(\gamma)}$ are shown for the first two orders of violation. The electron LIV parameter was set to $0$. The result for the first violation order was previously obtained in~\cite{Abdalla_2018}.}
    \label{fig:final:energy:ic:photon}
\end{figure}

We set $\delta_n = 0$ to isolate effects on the photon sector and solved Equation~\ref{eq:liv:ic} numerically. Figure~\ref{fig:final:energy:ic:photon} shows the LIV effect on the energy of the photon after the interaction. The plot shows how the energy of the photon after the interaction (final photon energy) depends on the energy of the photon before the interaction (initial photon energy) for different LIV coefficients. The energy of the photon after the interaction tends to the Thomson regime when the initial photon energy is larger than a value depending on the LIV coefficient, as shown in the plot. This can be verified as Equation~\ref{eq:final:energy:collision} and Equation~\ref{eq:liv:ic} leads to $E_\gamma' = E_\gamma$, which is the Thomson regime, in the high-energy limit ($\delta_n^{(\gamma)}E_\gamma^n \gg 1$). 

\begin{figure}[ht]
    \centering
    \includegraphics[width=0.8\textwidth]{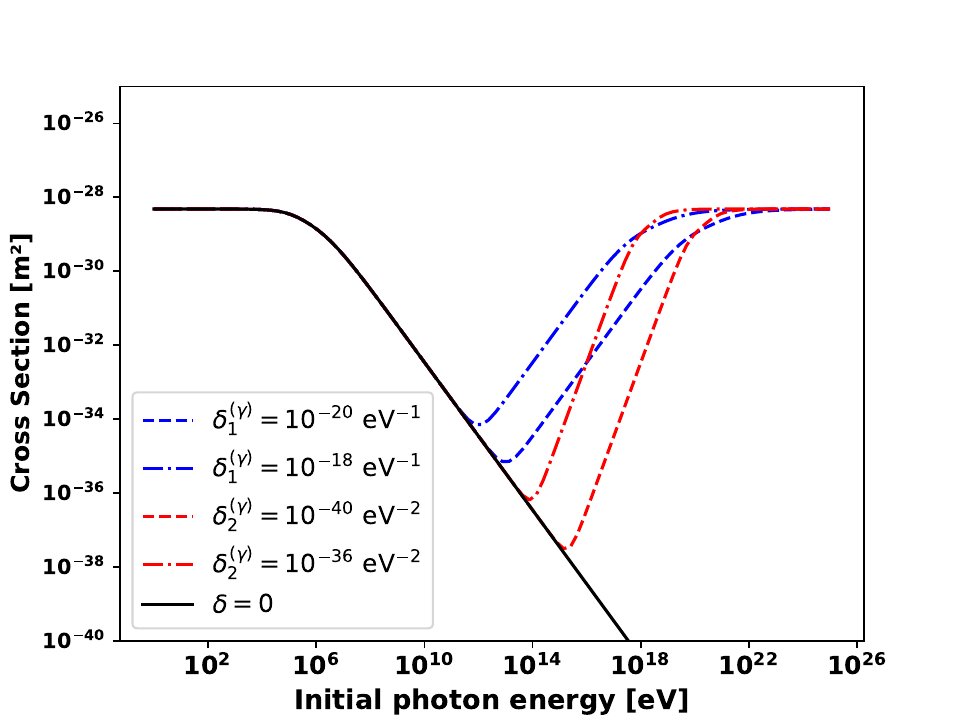}
    \caption{Klein-Nishina cross section for a photon-electron interaction. Different values of $\delta_n^{(\gamma)}$ are shown for the first two orders of violation. The electron LIV parameter was set to $0$. The result for the first violation order was previously obtained in~\cite{Abdalla_2018}.}
    \label{fig:cross:Section}
\end{figure}

Figure~\ref{fig:cross:Section} shows the evolution of the cross section with the initial energy of the photon, including LIV via Equations~\ref{eq:final:energy:collision} and~\ref{eq:liv:ic}. In our kinematic approach, LIV modifications to the photon dispersion relation lead to changes in the effective cross section, enhancing the interaction probability and effectively bypassing Klein-Nishina suppression until the Thomson limit is recovered.

\begin{figure}[ht]
    \centering
    \includegraphics[width=0.8\textwidth]{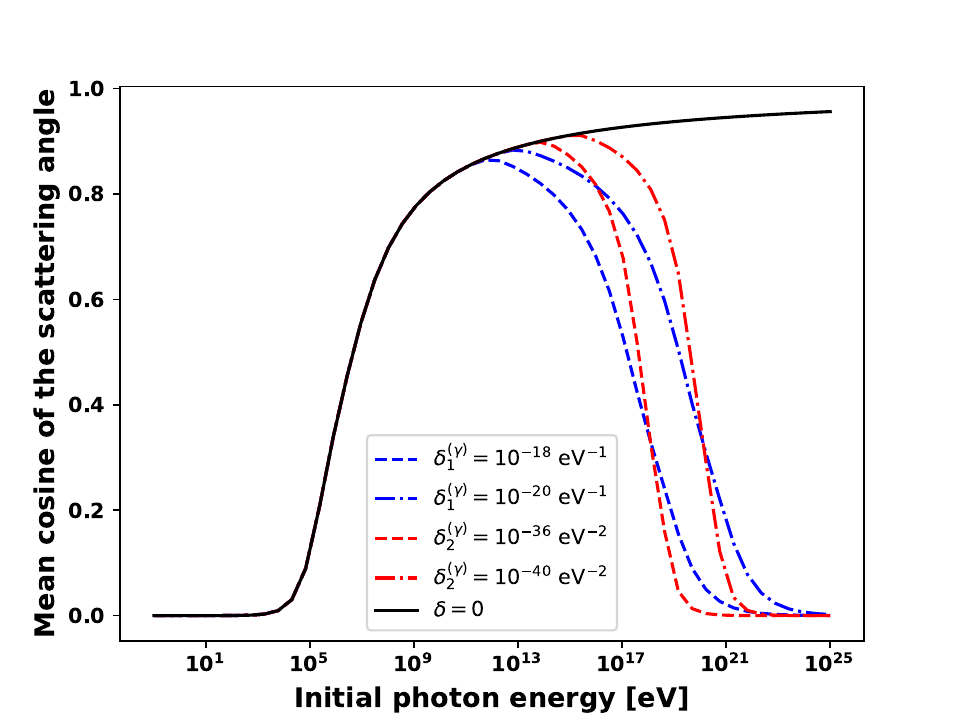}
    \caption{Average of the cosine of the scattering angle of photons as a function of the initial photon energy. Different values of $\delta_n^{(\gamma)}$ are shown for the first two orders of violation. The electron LIV parameter was set to $0$.}
    \label{fig:average}
\end{figure}

Figure~\ref{fig:average} shows the evolution of the scattering angle with the initial energy of the photon including LIV. A new transition to the Thomson regime is observed at high energies. 

Using the modified cross section and final energy of the photon, the emitted power under LIV assumption can be calculated via
\begin{equation}
P = \int \sigma_{\textrm{KN}} \left[E'_{\gamma} - E_{\gamma} \right]\eta(E_{\gamma})dE_{\gamma}\textrm{,}
\label{eq:power:def}
\end{equation}
where $\eta(E_\gamma)$ is the energy distribution of photons before the interaction. Within our framework, LIV increases both the final energy and cross section, thus amplifying the emitted power.

As mentioned in Section~\ref{sec:liv}, we do not introduce background timelike vectors or coefficients that would single out a preferred reference frame. Our analysis thus focuses exclusively on the modifications arising from the photon dispersion relation, without assuming any model-specific preferred-frame structure. Nonetheless, certain LIV formulations do predict the existence of a preferred frame, which could alter the mapping of scattering angles and change the average scattering angle obtained. As shown in Figure~\ref{fig:average}, angular corrections would be expected to appear at comparatively lower energies, while LIV modifications to the boost from the preferred frame to the charged particle frame are expected to become relevant only when $E^n \sim \delta_n^{-1}$. The energy scale at which the Thomson-like regime is recovered in our kinematic analysis coincides with the scale where LIV-induced boost effects would likely emerge. Hence, the actual existence and properties of this new regime remain uncertain until a complete analysis is performed within a specific LIV model that consistently includes both kinematic and dynamical effects. A full LIV-QED framework could introduce further corrections to the scattering amplitude and radiative processes. Nevertheless, the present treatment provides a controlled first step, highlighting the dominant kinematic effects within the inverse Compton process and offering a useful benchmark for comparison with more complete approaches.

\section{First-order Fermi acceleration including radiation process under LIV assumption}
\label{sec:firstorder}

First-order Fermi mechanism~\cite{Bell_1978,Krymskii_1977,blandford_1987} is an efficient acceleration process known to be working in several astrophysical sources. In a previous publication~\cite{Duarte_2024}, we investigated how LIV modifies the energy gain per scattering, the acceleration timescales, and the spectral indices of particles accelerated by the first-order Fermi mechanism. In this Section, we will merge the first-order Fermi with synchrotron and SSC energy losses, both including LIV within our kinematic framework.

It is important to note that, although we adopt a superluminal LIV framework, the restrictions imposed on the parameter $\delta_n$, as mentioned in Section~\ref{sec:liv}, ensures that particle velocities approach, but never exceed, the speed of light. Consequently, we do not consider other energy losses from processes that would otherwise become kinematically allowed, such as vacuum Cherenkov emission.

\subsection{Synchrotron radiation in first-order Fermi acceleration under LIV assumption}

The synchrotron loss process can be highly relevant in scenarios of astrophysical particle acceleration, mainly in two stages: initially, during the acceleration of the charged particles itself, where energy gains compete with radiative losses, and, subsequently, near the source, where the particles interact with the radiation fields. 

\subsubsection{Synchrotron: maximum energy of accelerated particles}
\label{subsec:max}

In Reference~\cite{Duarte_2024}, we calculated the energy gained by particles under first-order Fermi acceleration considering LIV. This modified energy gain, along with the power lost through synchrotron radiation, gives the following net gain $-b$ over time:
\begin{equation}
    b \cdot \tau_{avg} = -\left< \Delta E\right> + P_{sync} \cdot \tau_{avg} \rm{,}
\end{equation}
where $\tau_{avg}$ is the average crossing time for the particle traversing the shock wave. We consider the best-case scenario of Bohm diffusion~\cite{Bohm_1949}
\begin{equation}
    \tau_{avg} \approx \frac{5}{qBV}\frac{E}{\sqrt{1+\delta_nE^n}} \rm{.}
\end{equation}

In a scenario where synchrotron radiation is the dominant energy loss during shock wave acceleration, we can use Equation~\ref{eq:pot:sync} to determine the maximum energy a charged particle can gain through first-order Fermi acceleration~\cite{Kumar_2012} as the equilibrium between loss and energy gain, $\left< \Delta E \right> = P_{sync}\cdot\tau_{avg}$, which gives
\begin{equation}
    \frac{E_{\textrm{max}}^2}{\left[1 -\left(n+1 \right)\frac{\delta_n}{m^2}E^{n+2} \right]} \approx \frac{m^4}{B}\rm{.}
    \label{eq:equilibrium}
\end{equation}

\begin{figure}[ht]
    \centering
    \includegraphics[width=0.8\textwidth]{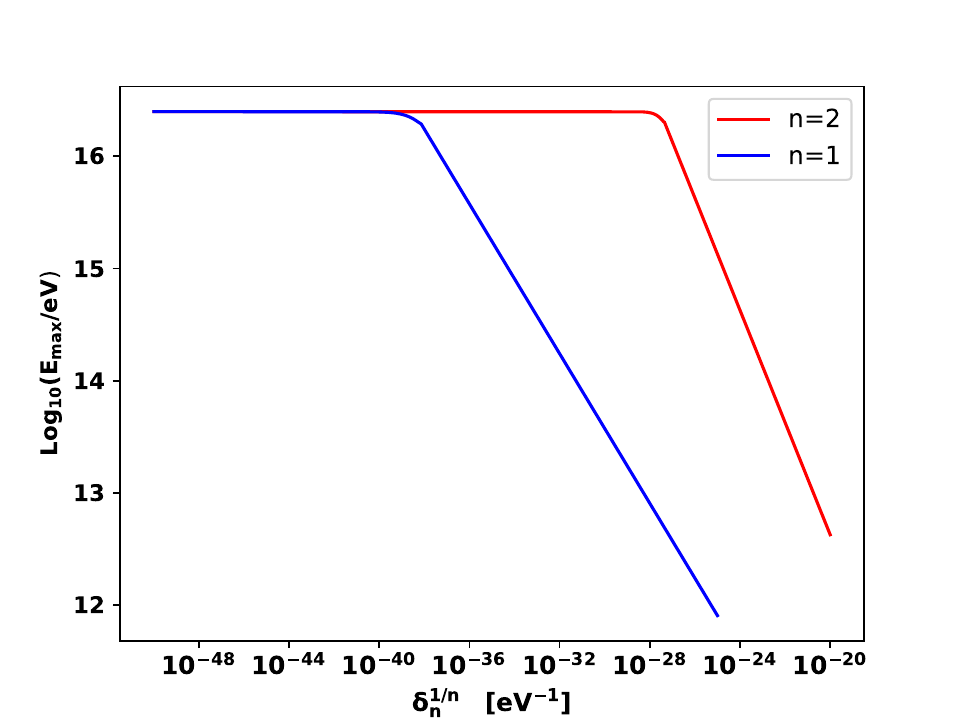}
    \caption{Maximum electron energy as a function of breaking parameter $\delta_n^{(e)}$. Electrons are accelerated via the first-order Fermi mechanism and are susceptible to synchrotron losses. A magnetic field of $B = 10^{-10} \ T$ was adopted. Both cases of $n=1$ and $n=2$ are shown.}
    \label{fig:max_elec}
\end{figure}

\begin{figure}[ht]
    \centering
    \includegraphics[width=0.8\textwidth]{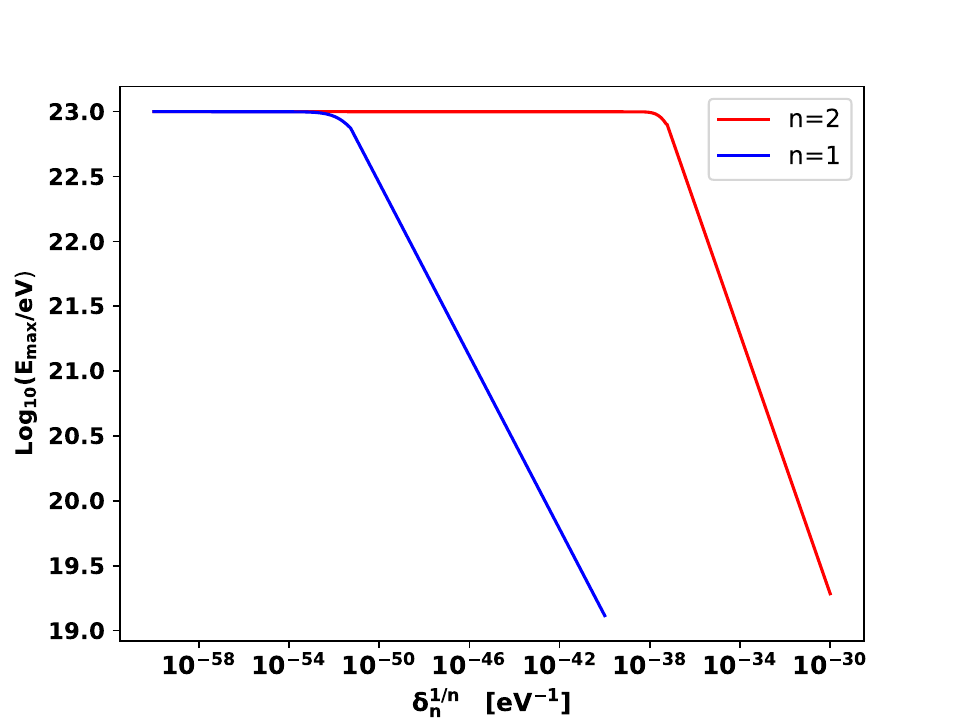}
    \caption{Maximum proton energy as a function of the breaking parameter $\delta_n^{(p)}$. Protons are accelerated via the first-order Fermi mechanism and are susceptible to synchrotron losses. A magnetic field of $B = 10^{-10} \ T$ was adopted. Both cases of $n=1$ and $n=2$ are shown.}
    \label{fig:max_prot}
\end{figure}

Figure~\ref{fig:max_elec} and Figure~\ref{fig:max_prot} show the solution to this equation as a function of $\delta_n$ for electrons and protons, respectively. A transition between the two regimes is very clear. In the regime of low $\delta_n$ values, the maximum energy is determined by the balance between energy gain and synchrotron losses for standard LI. As $\delta_n$ increases, kinematic LIV becomes more significant, and as $E^{n+2} \rightarrow \frac{m^2}{(n+1)\delta_n}$ the emitted power rises sharply, constraining the maximum attainable energy. This transition appears as an abrupt energy loss in the plots.

\subsubsection{Synchrotron: energy spectrum of charged particles and photons}
\label{subsub:sync}

We will use the one-zone model to calculate the energy spectrum of particles accelerated in the source~\cite{Gupta_2006}. This model considers particles traveling through an acceleration and a cooling (or energy-loss) region. The transport of particles can be calculated using the diffusion-loss equation to describe the number of particles $N$ within an energy range $E$ to $E+ dE$ at a given time
\begin{equation}
    \frac{dN}{dt} + \frac{d}{dE}\left[b\cdot N \right] = -\frac{N}{\tau_{scp}} + Q_{\mathrm{inj}} \rm{,}
    \label{eq:diffloss}
\end{equation}
where $b$ is the energy loss rate, in this case dominated by synchrotron radiation, $\tau_{\mathrm{esc}}$ is the escape timescale from the cooling region (here neglected), and $Q_{\mathrm{inj}}(E)$ is the injection spectrum from the acceleration site. We model $Q_{\mathrm{inj}}$ as a first-order Fermi spectrum with LIV and an exponential cutoff at $E_{\mathrm{max}}$ (derived in Subsection~\ref{subsec:max}).

\begin{figure}[ht]
    \centering
    \includegraphics[width=0.8\textwidth]{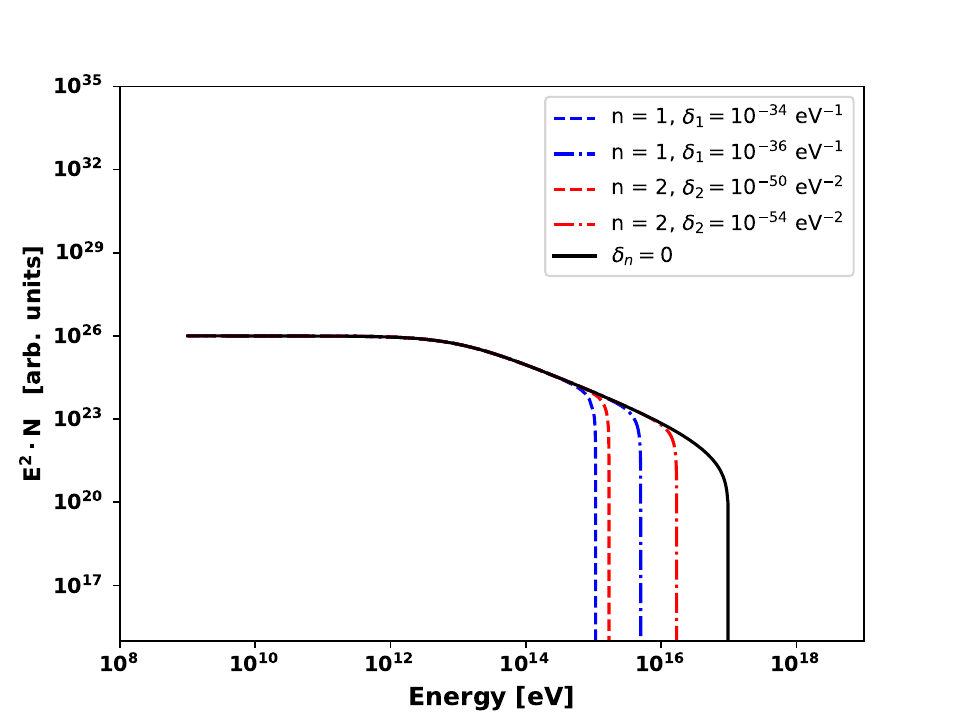}
    \caption{Electron's energy spectrum with an arbitrary normalization. The black solid line represents the Lorentz invariant scenario. The energy of the spectral index break was taken to be $E_b = 10^{13}\ \rm{eV}$. Different values of $\delta_p^{(e)}$ are shown for the studied orders of violation, $n=1$ and $n=2$. For each fixed value of $\delta_n$, the energy of the electron is restricted to $E^{n+2} < \frac{m^2}{(n+1)\delta_n}$.}
    \label{fig:electron:spec}
\end{figure}

\begin{figure}[ht]
    \centering
    \includegraphics[width=0.8\textwidth]{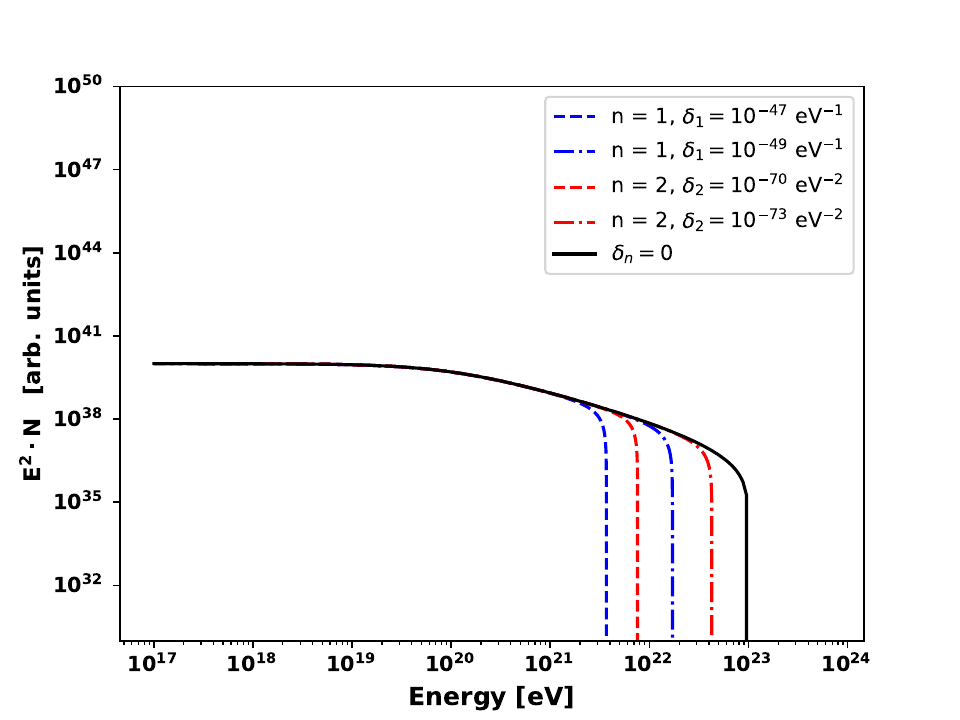}
    \caption{Proton's energy spectrum with an arbitrary normalization. The black solid line represents the Lorentz invariant scenario. The energy of the spectral index break was taken to be $E_b = 10^{20} \ \rm{eV}$. Different values of $\delta_n^{(p)}$  are shown for the studied orders of violation, $n=1$ and $n=2$. For each fixed value of $\delta_n$, the energy of the proton is restricted to $E^{n+2} < \frac{m^2}{(n+1)\delta_n}$.}
    \label{fig:proton:spec}
\end{figure}

Figures~\ref{fig:electron:spec} and~\ref{fig:proton:spec} show the energy spectrum of charged particles resulting from the one-zone model, including LIV for electrons and protons, respectively. A magnetic field of $B = 10^{-10} \ \textrm{T}$ was used. The abrupt decay in the spectrum is a direct consequence of the maximum energy derived in Section~\ref{subsec:max}. Although the emitted power formally diverges at higher energies, in our analysis we restrict ourselves to values of energy and $\delta_n$ for which the Lorentz factor remains physically valid. In this sense, the behavior observed in the plots reflects the trend expected when kinematic LIV is taken into account. To obtain quantitative bounds from this behavior, however, a complete dynamical treatment would be required given that the divergences and unphysical results could be modified in a fully consistent framework.

The energy spectrum of photons produced by the accelerated particles via synchrotron emission can be modeled by~\cite{Ghisellini_2013}
\begin{equation}
    N_{\gamma}(E_{\gamma}) =\frac{D}{4\pi}\int_{E_{min}}^{E_{max}} P(E) N_{p}(E) dE \rm{,}
    \label{eq:emission}
\end{equation}
where $P$ is the synchrotron power, $N_p$ is the particle's energy spectrum, and $D$ is the volume of the source. As previously discussed, we assume all photons are emitted with the same energy at the peak frequency, resulting in
\begin{equation}
    N_{\gamma}(E_{\gamma}) \approx 10^{14} \frac{DB}{m}N_{p}(E) E \left[\frac{1 -\left(n+1 \right)\frac{\delta_n}{m^2}E^{n+2}}{1+(n+1)^2\frac{\delta_n}{m^2}E^{n+2}} \right] \rm{,}
    \label{eq:photon:emission}
\end{equation}
where $E$ is the energy of the particle that emits a photon with energy $E_{\gamma}$ according to Equation~\ref{eq:photon:sync}, and $N_p(E)$ is the spectrum calculated using Equation~\ref{eq:diffloss}.

\begin{figure}[ht]
    \centering
    \includegraphics[width=0.8\textwidth]{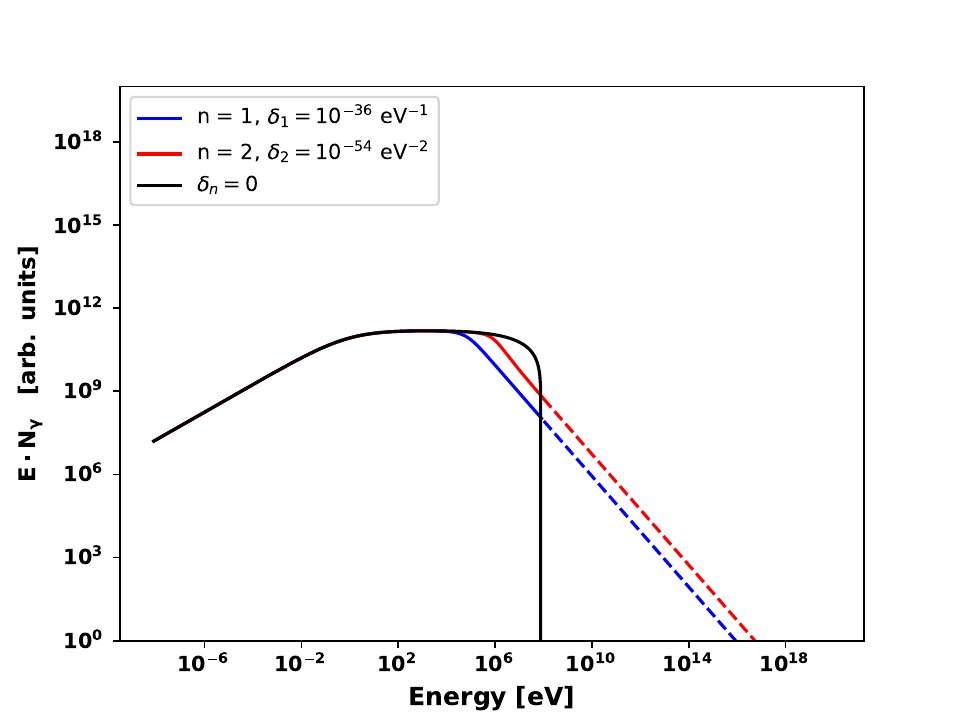}
    \caption{Photon's energy spectrum emitted via synchrotron emission by a population of electrons with an arbitrary normalization. The energy of the spectral index break was taken to be $E_b = 10^{13} \ \rm{eV}$. A magnetic field of $B = 10^{-10} \ T$ was adopted. The black solid line represents the Lorentz invariant scenario. The dashed lines indicate the energy region at which energy conservation is violated. Different values of $\delta_n^{(e)}$ are shown for the studied orders of violation, $n=1$ and $n=2$. For each fixed value of $\delta_n$, the energy of the electron emitting radiation is restricted to $E^{n+2} < \frac{m^2}{(n+1)\delta_n}$.}
    \label{fig:electron:photon}
\end{figure}

\begin{figure}[ht]
    \centering
    \includegraphics[width=0.8\textwidth]{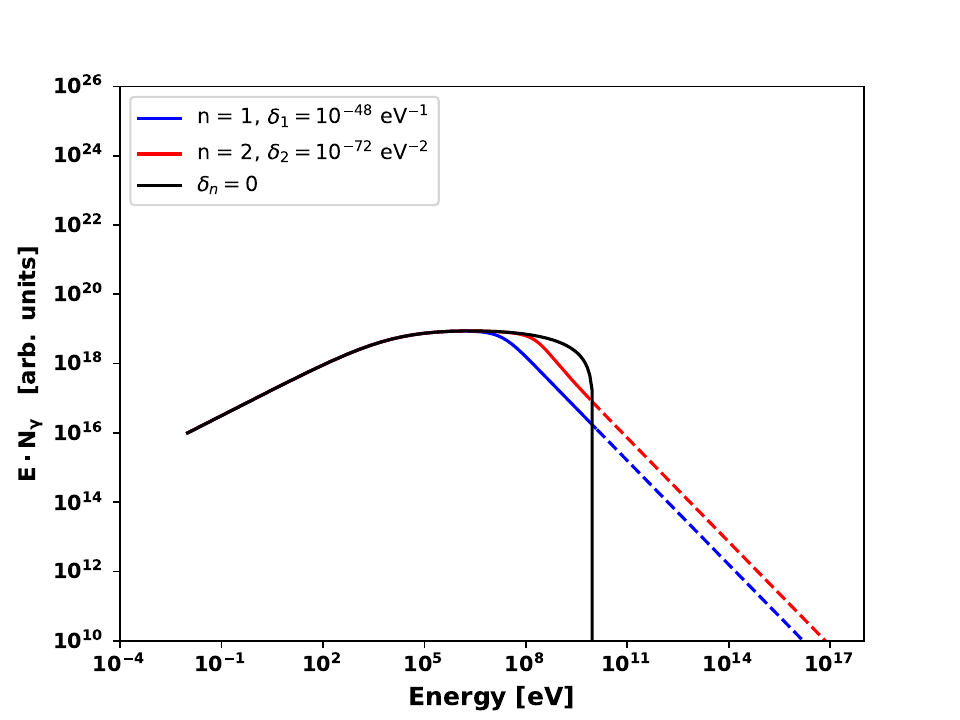}
    \caption{Photon's energy spectrum emitted via synchrotron emission by a population of protons with an arbitrary normalization. The energy of the spectral index break was taken to be $E_b = 10^{20}\ \rm{eV}$. A magnetic field of $B = 10^{-10} \ T$ was adopted. The black solid line represents the Lorentz invariant scenario. The dashed lines indicate the energy region at which energy conservation is violated. Different values of $\delta_n^{(p)}$ are shown for the studied orders of violation, $n=1$ and $n=2$. For each fixed value of $\delta_n$, the energy of the electron emitting radiation is restricted to $E^{n+2} < \frac{m^2}{(n+1)\delta_n}$.}
    \label{fig:proton:photon}
\end{figure}

Figures~\ref{fig:electron:photon} and~\ref{fig:proton:photon} show the resulting energy spectrum for electrons and protons, respectively. In these calculations, we considered a source with a size of $1\hspace{0.3em}\rm{pc}$.

LIV leads to the emergence of a distinct high-energy emission, where the final photon energy exhibits rapid growth near the divergence threshold - $E^{n+2} \rightarrow \frac{m^2}{(n+1)\delta_n}$ - expected from Equation~\ref{eq:photon:sync}. Notably, this region contributes a non-negligible flux, consistent with the increasing emitted power derived in Equation~\ref{eq:pot:sync}. However, it is important to note that the energy of the emitted photon exceeds that of the electron or proton responsible for its production, thus violating energy conservation. The energy range where this violation occurs is indicated by a dashed line in Figures~\ref{fig:electron:photon} and~\ref{fig:proton:photon}.

As noted earlier, these results were obtained solely by incorporating the modified dispersion relation, with the goal of isolating the dominant kinematical LIV effects on synchrotron emission. A full analysis of this process within a dynamical LIV–QED framework may confirm or modify these conclusions, and we leave such an investigation for future work. Should the current trend persist, LIV not only introduces an unanticipated energy domain in synchrotron emission but could also alter SSC spectra through the inclusion of higher-energy photons in the radiation background, resulting in emissions at previously unexpected energies and potentially altering the predicted spectra. These modifications could serve as a means to constrain LIV, especially when combined with propagation effects, since additional processes such as photon splitting \cite{Kaufhold:2005vj, huerta} are expected to become relevant in this extended energy range of emission.

\subsection{Synchrotron self-Compton in first-order Fermi acceleration under LIV assumption}

In non-thermal astrophysical spectra, the population of higher-energy photons is commonly attributed to the inverse Compton scattering. In this process, background photons near the source gain energy from charged particles in the acceleration region.

In regions with a high concentration of electrons and strong magnetic fields, the SSC model can become dominant~\cite{Nalewajko:2017zbt,Kusunose:2008dd,Ghisellini_2013}. In this framework, the seed photons originate from the synchrotron radiation produced by the particle population itself. This study will focus on this type of process, where the photon density in Equation~\ref{eq:power:def} is determined by the synchrotron emission presented in Equation~\ref{eq:photon:emission}, while the cross section and final photon energy are the same as discussed in Subsection~\ref{subsec:ic:theory}. Notice that $\delta_n^{(\gamma)}$ will not affect the photon distribution; only the inverse Compton scattering will be affected by this parameter. Again, the parameter $\delta_n$ will be set to $0$ in order to focus exclusively on the behavior of LIV in photons.

\subsubsection{Synchrotron self-Compton: energy spectrum of charged particles and photons}

\begin{figure}[ht]
    \centering
    \includegraphics[width=0.8\textwidth]{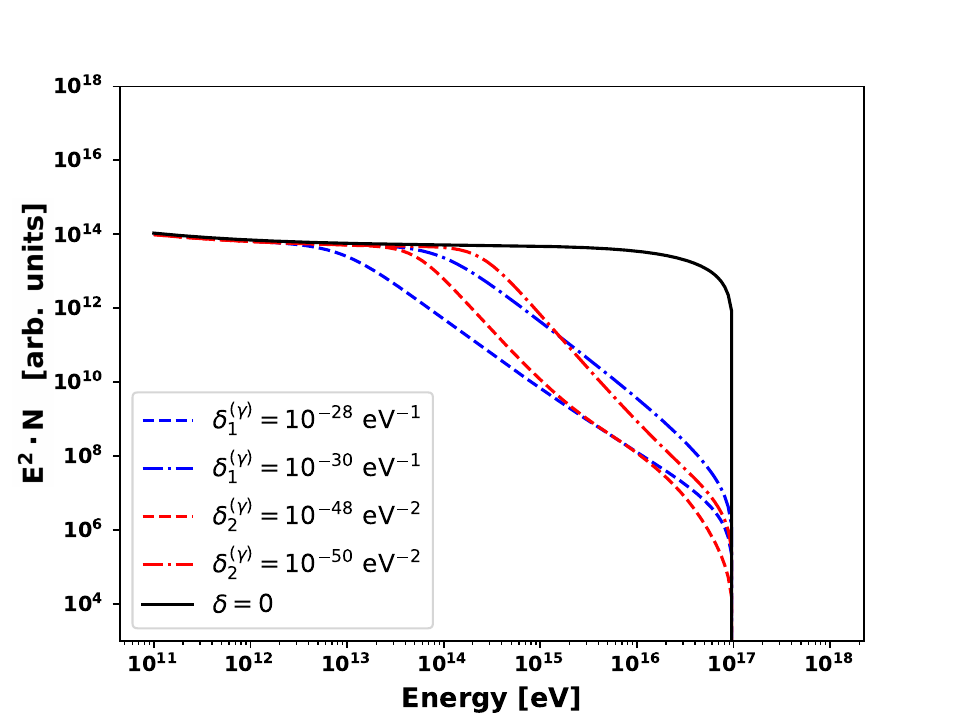}
    \caption{Electron's energy spectrum resulting from the one-zone model with an arbitrary normalization. The electron breaking parameter was set to $0$. The energy of the spectral index break was taken to be $E_b = 10^{13}\ \rm{eV}$. The black solid line represents the Lorentz invariant scenario. Different values of $\delta_n^{(\gamma)}$ are shown for the studied orders of violation, $n=1$ and $n=2$.}
    \label{fig:electron:ic}
\end{figure}

\begin{figure}[ht]
    \centering
    \includegraphics[width=0.8\textwidth]{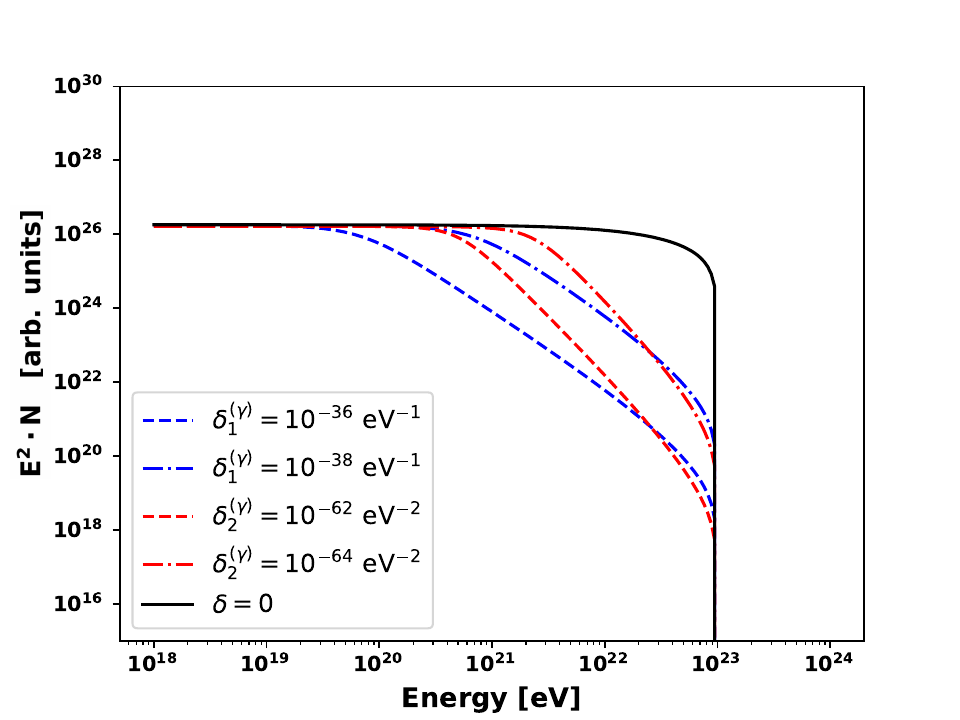}
    \caption{Proton's energy spectrum resulting from the one-zone model with an arbitrary normalization. The proton breaking parameter was set to $0$. The energy of the spectral index break was taken to be $E_b = 10^{20}\ \rm{eV}$. The black solid line represents the Lorentz invariant scenario. The dashed lines indicate the energy region at which energy conservation is violated. Different values of $\delta_n^{(\gamma)}$ are shown for the studied orders of violation, $n=1$ and $n=2$.}
    \label{fig:proton:ic:nodelta}
\end{figure}

In this section, we repeat the calculations in Subsection~\ref{subsub:sync}, now considering SSC as the dominant energy-loss mechanism. Figures~\ref{fig:electron:ic} and~\ref{fig:proton:ic:nodelta} show the energy spectrum of electrons and protons, respectively. A suppression of the flux with increasing energy is seen in both cases. The suppression caused by LIV is well below in energy and has less abrupt fall when compared to the cutoff in SSC under LI assumption. The LIV suppression is caused by two effects, the final energy in the interaction and the cross section, which tend to increase rapidly when LIV is considered, inducing a growth in the emitted power.

\begin{figure}[ht]
    \centering
    \includegraphics[width=0.8\textwidth]{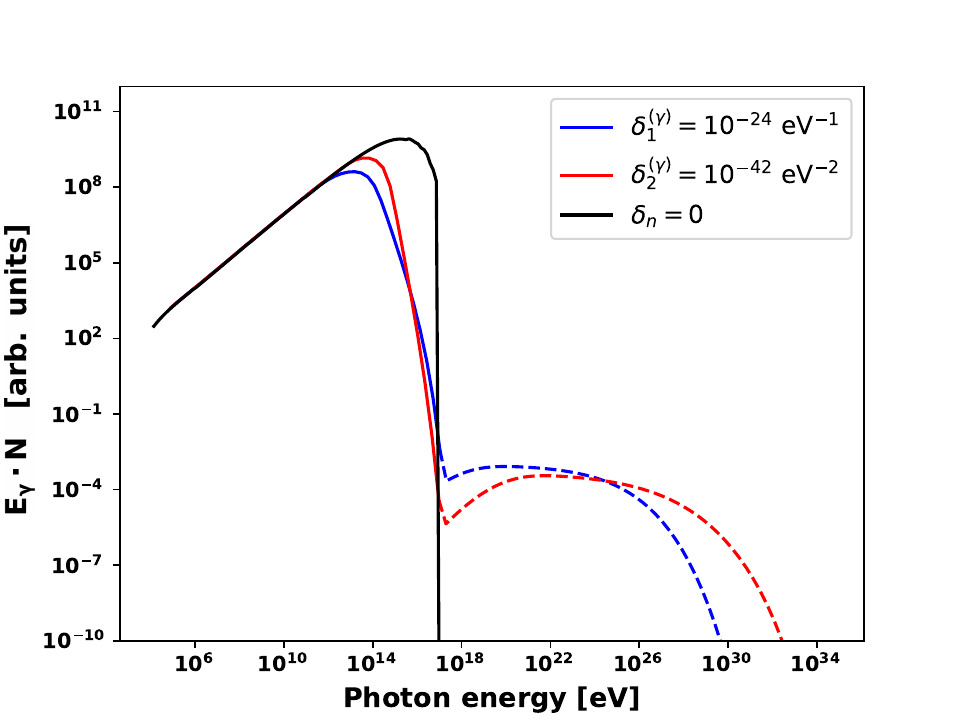}
    \caption{Photon's energy spectrum emitted via synchrotron self-Compton by a population of electrons with an arbitrary normalization. Inverse Compton radiation was considered the dominant loss process. The energy of the spectral index break was taken to be $E_b = 10^{11}\ \rm{eV}$. The black solid line represents the Lorentz invariant scenario. The dashed lines indicate the energy region at which energy conservation is violated. Different values of $\delta_n^{(\gamma)}$ are shown for the studied orders of violation, $n=1$ and $n=2$.}
    \label{fig:electron:photon:ic}
\end{figure}

\begin{figure}[ht]
    \centering
    \includegraphics[width=0.8\textwidth]{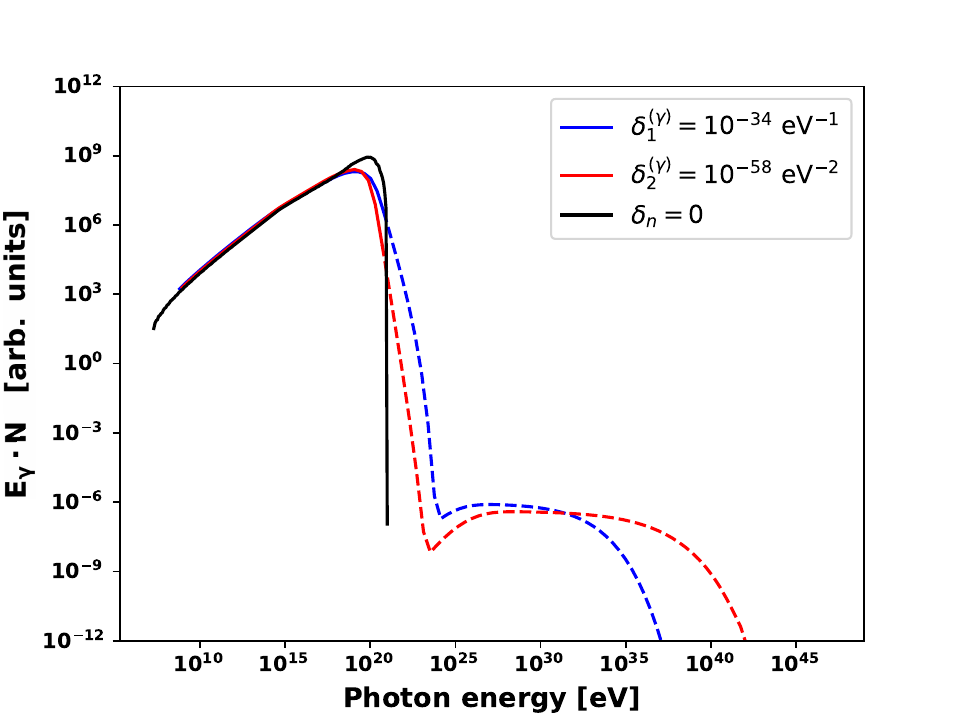}
    \caption{Photon's energy spectrum emitted via synchrotron self-Compton by a population of protons with an arbitrary normalization. Inverse Compton radiation was considered the dominant loss process. The energy of the spectral index break was taken to be $E_b = 10^{20}\ \rm{eV}$. The black solid line represents the Lorentz invariant scenario. Different values of $\delta_n^{(\gamma)}$ are shown for the studied orders of violation, $n=1$ and $n=2$.}
    \label{fig:proton:photon:ic}
\end{figure}

Following the same arguments of the previous section, we also calculated the energy spectrum of photons produced in the SSC interaction of the charged particles. Figure~\ref{fig:electron:photon:ic} and~\ref{fig:proton:photon:ic} show the energy spectrum of photons resulting from the interaction of electrons and protons, respectively. A magnetic field of $B = 10^{-10} \ \textrm{T}$ was considered.

\begin{figure}[ht]
    \centering
    \includegraphics[width=0.8\textwidth]{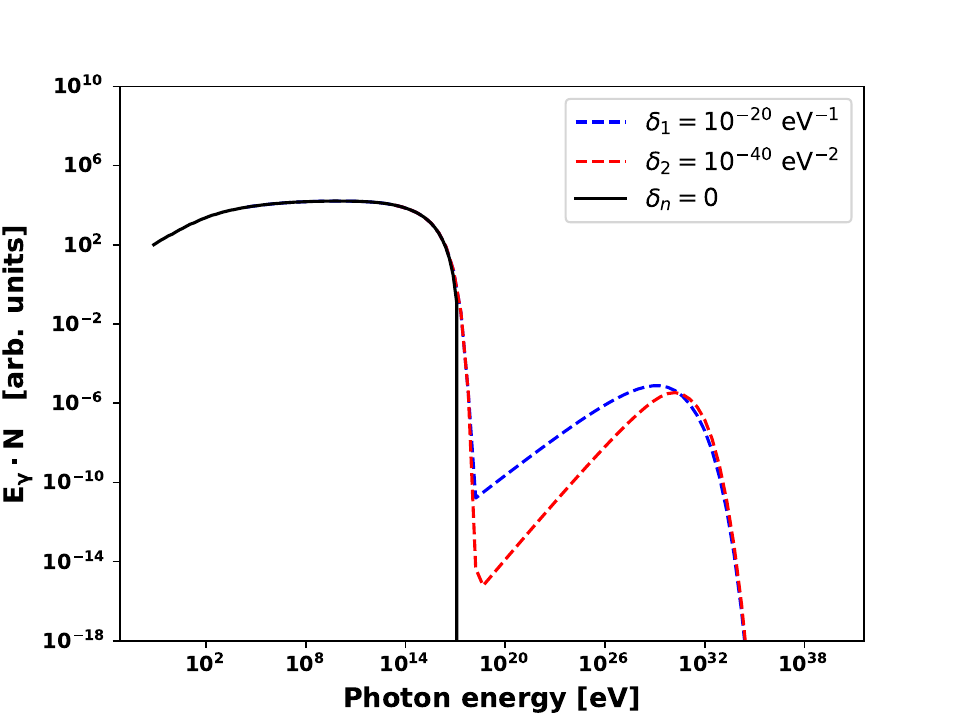}
    \caption{Photon's energy spectrum emitted via synchrotron self-Compton by a population of electrons with an arbitrary normalization. Synchrotron emission was considered the dominant loss process. The energy of the spectral index break was taken to be $E_b = 10^{11}\ \rm{eV}$. The black solid line represents the Lorentz invariant scenario. The dashed lines indicate the energy region at which energy conservation is violated. Different values of $\delta_n^{(\gamma)}$ are shown for the studied orders of violation, $n=1$ and $n=2$.}
    \label{fig:electron:photon:ic_sync}
\end{figure}

The photons emitted by protons and electrons exhibit similar behavior, differing primarily in their LIV parameters due to the proton's higher energy. As shown in Figures~\ref{fig:electron:photon:ic} and~\ref{fig:proton:photon:ic}, a new emission region emerges at higher energies due to LIV. The peak height depends on the magnetic field strength, with stronger fields producing more energetic background photons, enhancing LIV effects. The charged-particle cutoff energy also influences emission in this region — lower cutoffs reduce its prominence. We assume SSC losses dominate, significantly suppressing the particle energy spectrum (Figures~\ref{fig:electron:ic}, \ref{fig:proton:ic:nodelta}). Therefore, competition arises between Fermi-accelerated spectral shaping and emission effects: a stronger violation increases emission but suppresses the electron/proton population responsible for high-energy photons. This suppression also shifts the standard SSC peak to lower energies. In contrast, for systems dominated by synchrotron losses, only emission modifications persist without spectral suppression, as shown in Figure~\ref{fig:electron:photon:ic_sync}.

The same discussion presented in Subsection~\ref{subsub:sync} regarding the emergence of new emission regimes and their impact on the particle spectrum also applies here. We highlight the exploratory character of these results and outline a possible strategy for incorporating LIV effects into the process.

\section{Conclusion}
\label{sec:conclusion}

In this work, we examined the influence of Lorentz invariance violation (LIV) on the production and emission of high-energy charged particles and photons in astrophysical environments. By modifying synchrotron and inverse Compton processes within the LIV framework, we derived the resulting changes in radiated power, photon energy, and energy spectra of accelerated particles. Our results demonstrate that LIV induces rapid growth in the Lorentz factor at high energies, leading to divergences in emission characteristics and suggesting a breakdown of the perturbative formalism, requiring new theoretical treatments at extreme energies. The model employed here is purely kinematical; a fully dynamical analysis could introduce additional corrections to both cross sections and emission spectra, and may alter some of the behaviors reported in this work.

Incorporating these effects into first-order Fermi acceleration models and synchrotron self-Compton scenarios, we showed that LIV modifies both particle spectra and the resulting multi-wavelength photon emission. In particular, LIV predicts the emergence of distinct high-energy components in the photon spectrum, while suppressing the energy spectrum of charged particles. These features of the photon emission were evaluated in astrophysical study cases, including the GeV–TeV emission from Markarian 501 and ultra-high-energy photons from a population of BL Lac objects (Appendix~\ref{sec:blazar}).

Looking forward, once a consistent dynamical formulation of LIV is established to ensure the validity of these behaviors, a multi-messenger analysis becomes a natural extension. LIV could simultaneously impact the photon sector and the spectrum of charged particles accelerated through first-order Fermi mechanisms, establishing correlated signatures across different messengers.

While current observational data, such as the upper limits from the Pierre Auger Observatory, do not yet constrain the explored LIV parameters, our results demonstrate that astrophysical photon spectra provide a promising avenue for probing violations of Lorentz symmetry, especially if the results here presented are maintained in a fully consistent dynamical formulation. Future multi-messenger observations, with improved sensitivity to both charged particle and photon fluxes, may offer critical tests of fundamental symmetries at the highest energies.
%
\section*{Acknowledgments}

The authors are supported by the S\~{a}o Paulo Research Foundation (FAPESP) through grant number 2021/01089-1. VdS is supported by CNPq through grant number 308837/2023-1. MDF is supported by CAPES through grant number 88887.684414/2022-00. The authors acknowledge the National Laboratory for Scientific Computing (LNCC/MCTI,  Brazil) for providing HPC resources for the SDumont supercomputer (http://sdumont.lncc.br).

\bibliographystyle{plainnat}
\bibliography{references}


\appendix
\section{Illustrative application to blazars}
\label{sec:blazar}

In the main text, we analyzed idealized cases of acceleration, emission, and related processes in order to isolate the effects introduced by Lorentz invariance violation (LIV). As an illustrative extension, we present here a simplified application to blazars, focusing on the case of Markarian~501 and, more generally, on the BL Lac population. The purpose of this appendix is not to derive quantitative constraints on the LIV coefficients, but rather to highlight possible observational signatures in scenarios where dynamical effects and the existence of preferred frames do not alter the trends previously reported.

\begin{figure}[ht]
    \centering
    \includegraphics[width=0.8\textwidth]{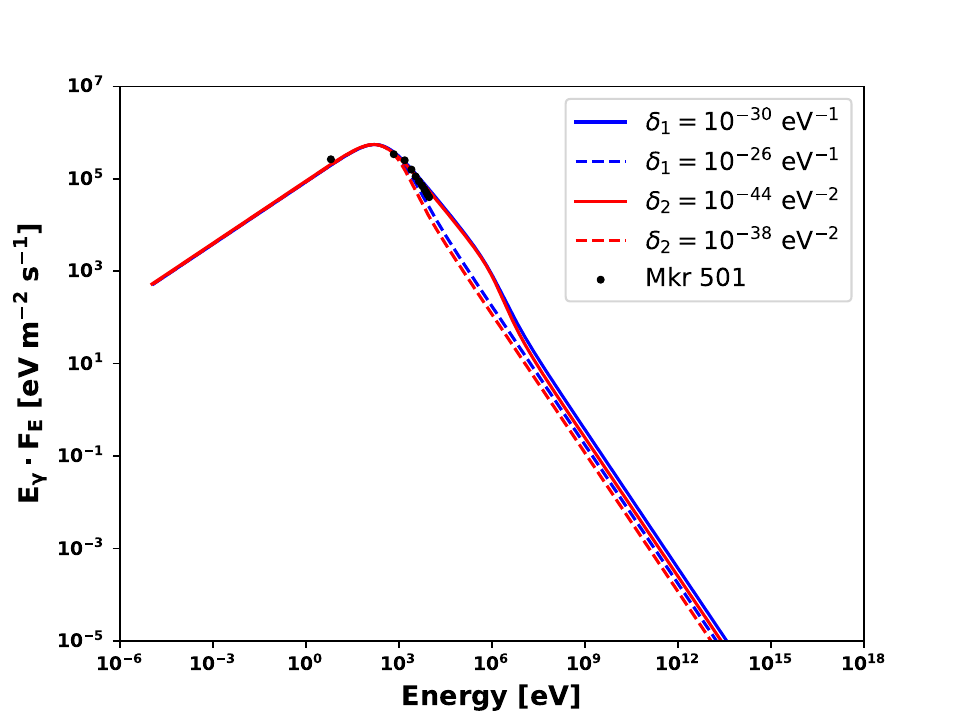}
    \caption{Photon's energy spectrum resulting from the synchrotron emission of Markarian 501. The data was taken from~\cite{Albert_2020} and only the synchrotron peak is shown. Different values of $\delta_n^{(e)}$ are shown for both violation orders, $n=1$ and $n=2$.}
    \label{fig:mkr_sync}
\end{figure}

\begin{figure}[ht]
    \centering
    \includegraphics[width=0.8\textwidth]{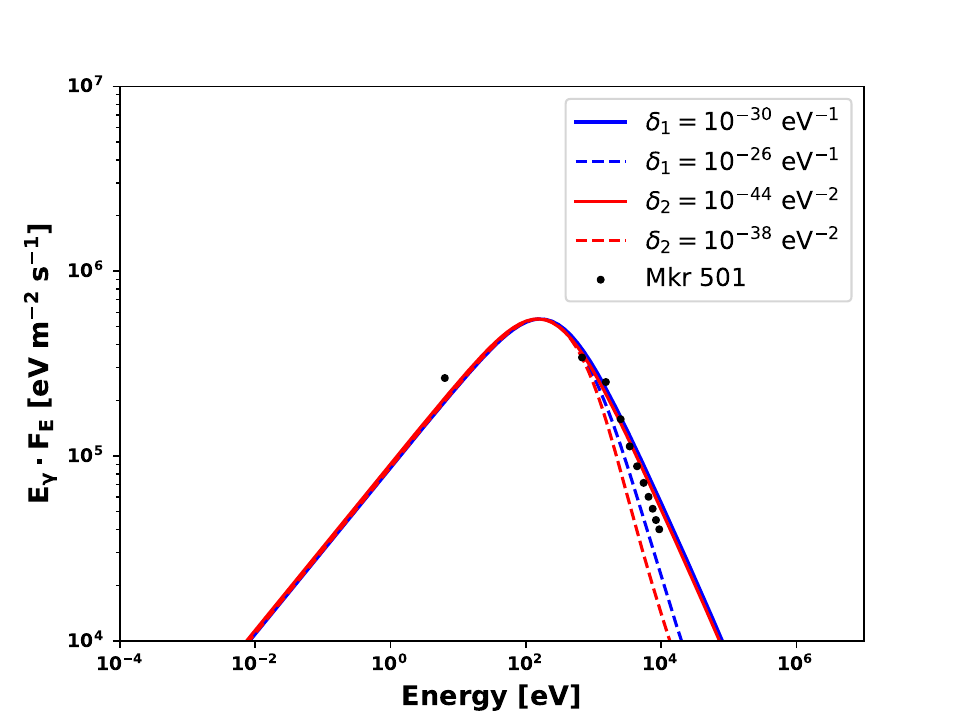}
    \caption{Zoom on the photon's energy spectrum resulting from the synchrotron emission of Markarian 501. The data was taken from~\cite{Albert_2020} and only the synchrotron peak is shown. Different values of $\delta_n^{(e)}$ are shown for both violation orders, $n=1$ and $n=2$.}
    \label{fig:mkr_sync_zoom}
\end{figure}

\begin{figure}[ht]
    \centering
    \includegraphics[width=0.8\textwidth]{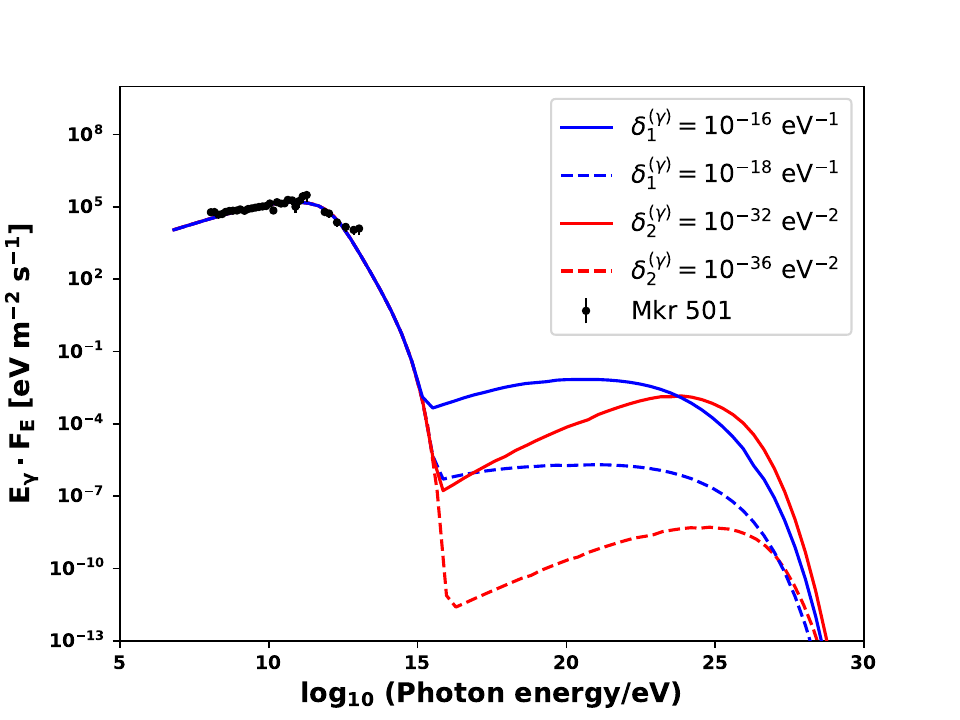}
    \caption{Photon's energy spectrum resulting from the SSC emission of Markarian 501. The data was taken from~\cite{Albert_2020} and only the SSC peak is shown. Different values of $\delta_n^{(\gamma)}$ are shown for both violation orders, $n=1$ and $n=2$.}
    \label{fig:mkr}
\end{figure}

\begin{figure}[ht]
    \centering
    \includegraphics[width=0.8\textwidth]{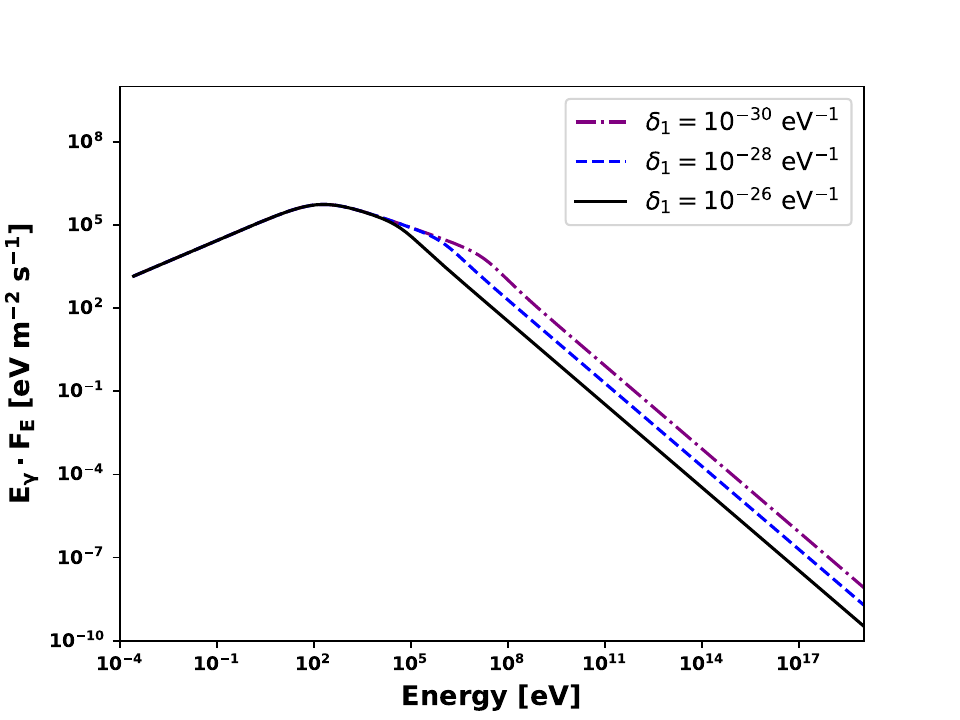}
    \caption{Photon's energy spectrum resulting from the synchrotron emission of a BL Lac object. A cut-off of $10^{14}\ \rm{eV}$ was adopted. The first order of violation, $n=1$, is shown for different breaking parameters in electrons.}
    \label{fig:bllac:spectrum1_sync}
\end{figure}

\begin{figure}[ht]
    \centering
    \includegraphics[width=0.8\textwidth]{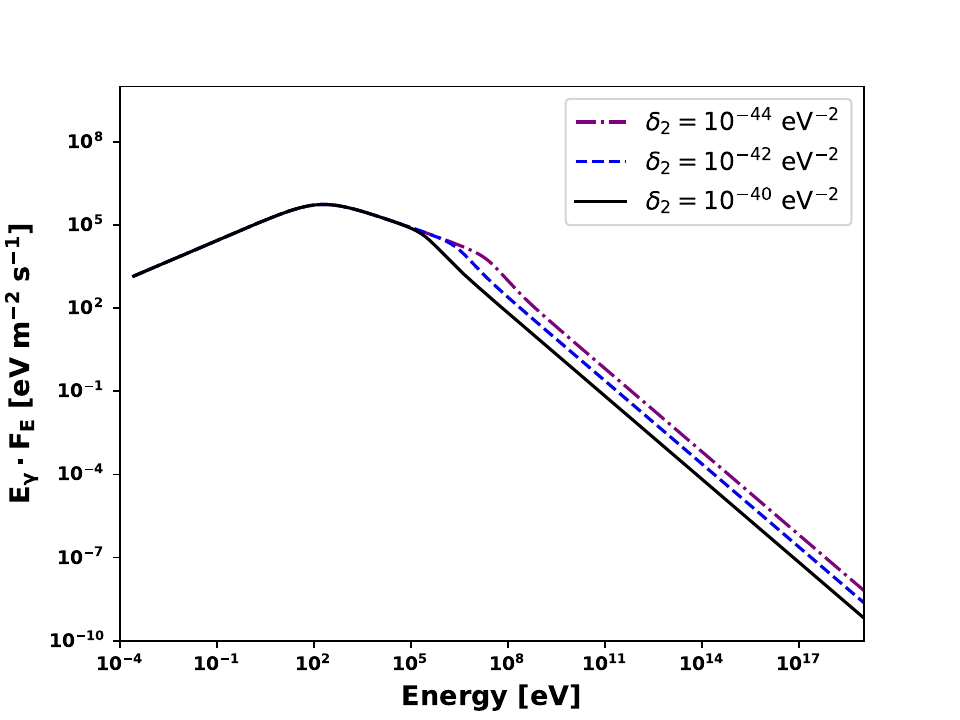}
    \caption{Photon's energy spectrum resulting from the synchrotron emission of a BL Lac object. A cut-off of $10^{14}\ \rm{eV}$ was adopted. The second order of violation, $n=2$, is shown for different breaking parameters in electrons.}
    \label{fig:bllac:spectrum2_sync}
\end{figure}

\begin{figure}[ht]
    \centering
    \includegraphics[width=0.8\textwidth]{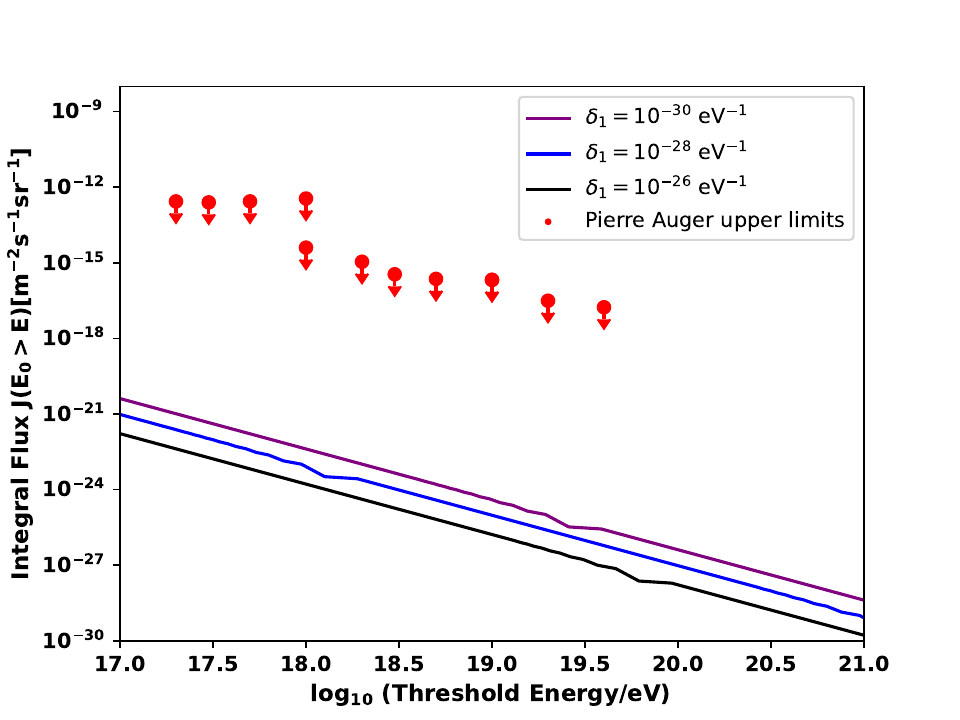}
    \caption{Integral photon flux for energies beyond $10^{17}\ \rm{eV}$ for a violation of order $n=1$ adopting different values of $\delta_1^{(e)}$. Red dots are the current upper limits set by the Pierre Auger Observatory.}
    \label{fig:bllac:limits1_sync}
\end{figure}

\begin{figure}[ht]
    \centering
    \includegraphics[width=0.8\textwidth]{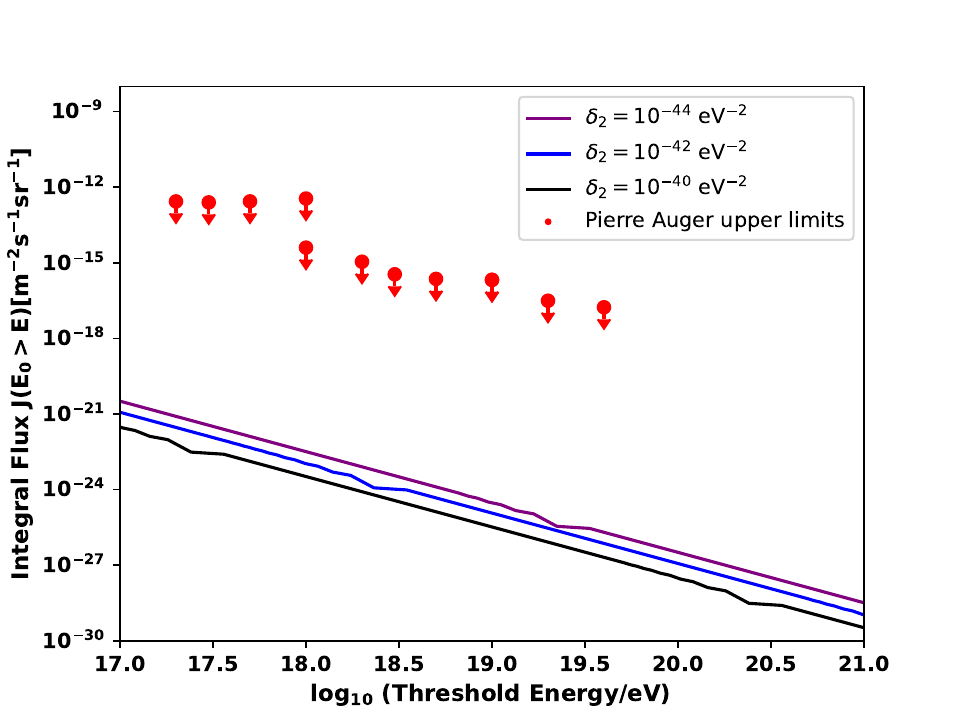}
    \caption{Integral photon flux for energies beyond $10^{17}\ \rm{eV}$ for a violation of order $n=2$ adopting different values of $\delta_2^{(e)}$. Red dots are the current upper limits set by the Pierre Auger Observatory.}
    \label{fig:bllac:limits2_sync}
\end{figure}

\begin{figure}[ht]
    \centering
    \includegraphics[width=0.8\textwidth]{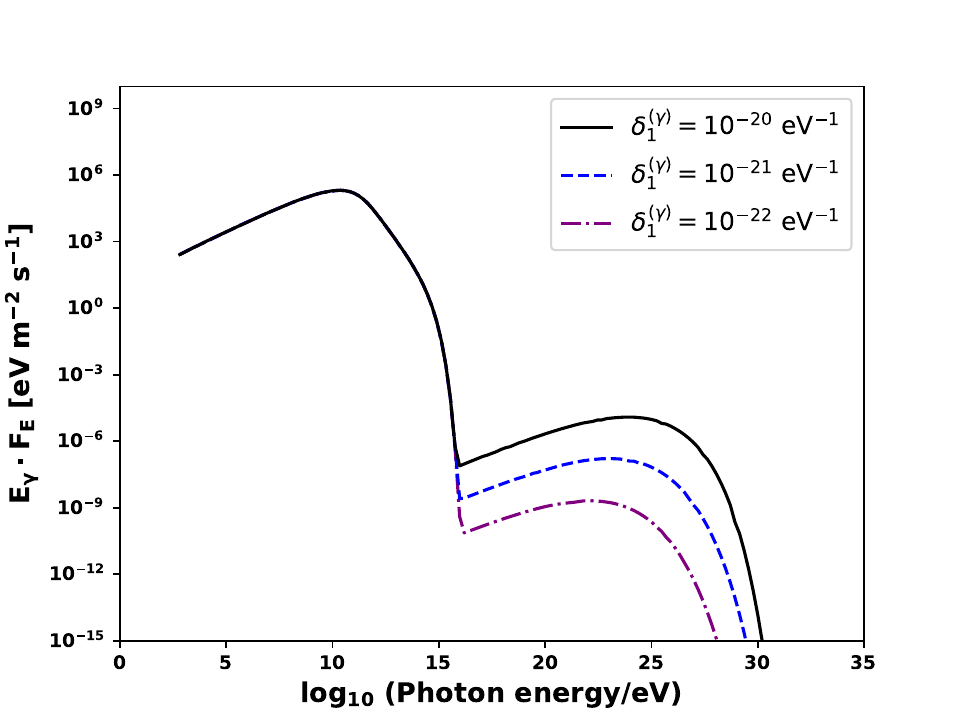}
    \caption{Photon's energy spectrum resulting from the SSC emission of a BL Lac object. A cut-off of $10^{14}\ \rm{eV}$ was adopted. The first order of violation, $n=1$, is shown for different breaking parameters in photons.}
    \label{fig:bllac:spectrum1}
\end{figure}

\begin{figure}[ht]
    \centering
    \includegraphics[width=0.8\textwidth]{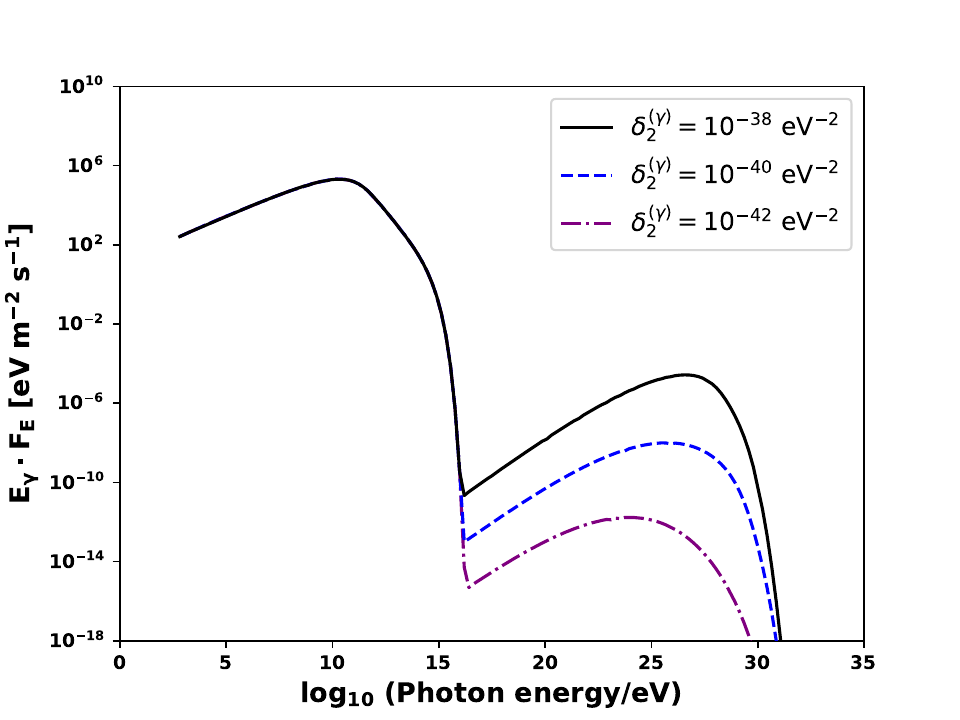}
    \caption{Photon's energy spectrum resulting from the SSC emission of a BL Lac object. A cut-off of $10^{14}\ \rm{eV}$ was adopted. The second order of violation, $n=2$, is shown for different breaking parameters in photons.}
    \label{fig:bllac:spectrum2}
\end{figure}

\begin{figure}[ht]
    \centering
    \includegraphics[width=0.8\textwidth]{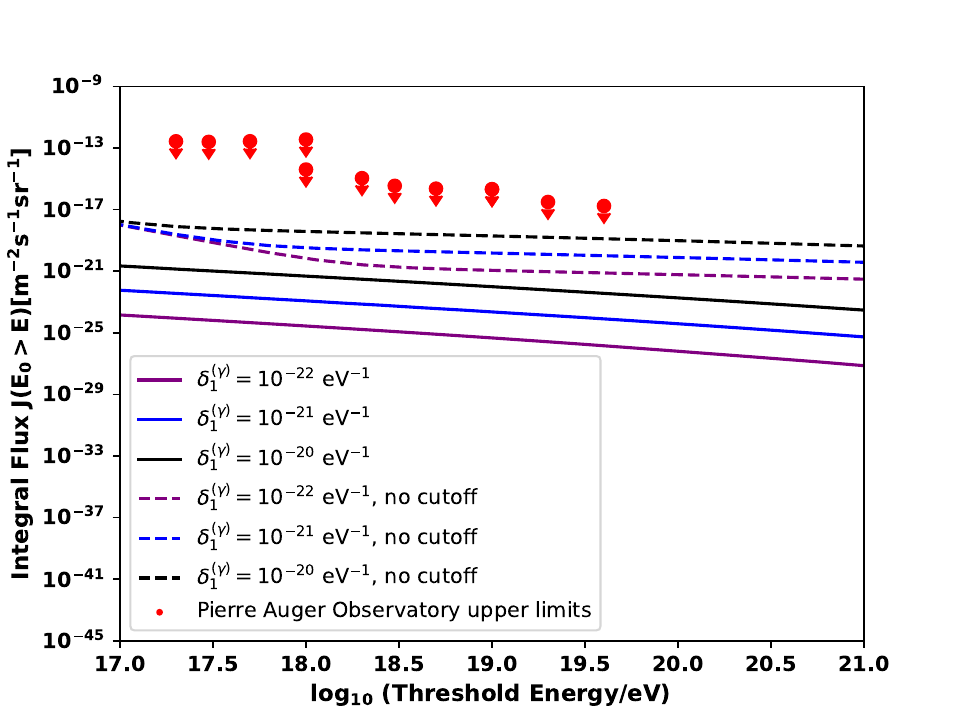}
    \caption{Integral photon flux for energies beyond $10^{17}\ \rm{eV}$ for a violation of order $n=1$ adopting different values of $\delta_1^{(\gamma)}$. The solid curves are the integral of photon fluxes for a population of $10^3$ BL Lac sources with a cut-off of $10^{14}\ \rm{eV}$. Dashed lines represent the same emission without cut-offs. Red dots are the current upper limits set by the Pierre Auger Observatory.}
    \label{fig:bllac:limits1}
\end{figure}

\begin{figure}[ht]
    \centering
    \includegraphics[width=0.8\textwidth]{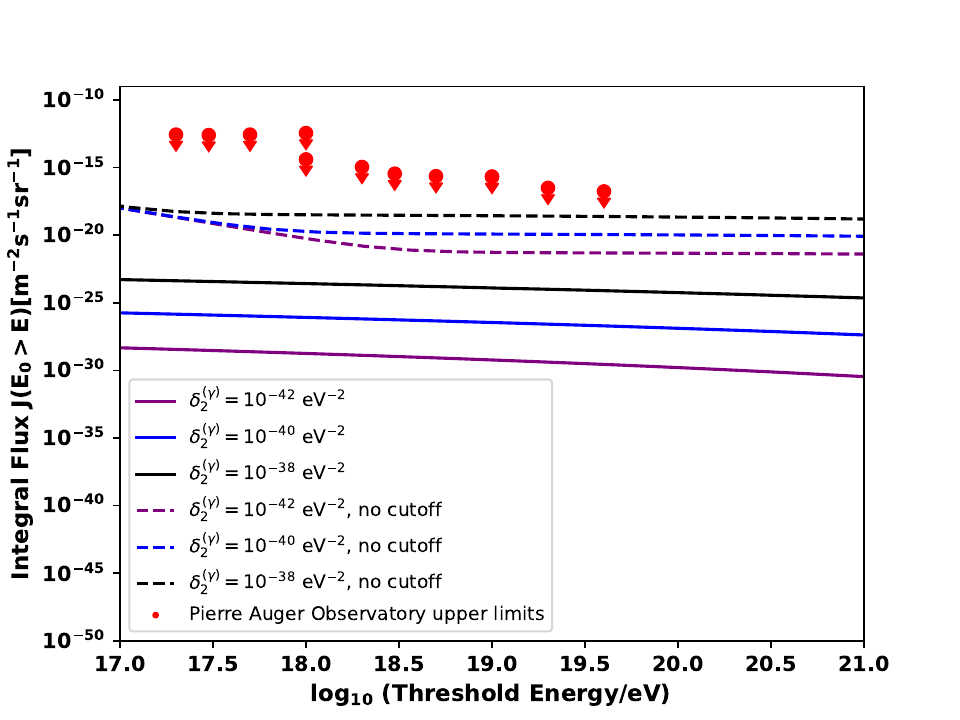}
    \caption{Integral photon flux for energies beyond $10^{17}\ \rm{eV}$ for a violation of order $n=2$ adopting different values of $\delta_2^{(\gamma)}$. The solid curves are the integral of photon fluxes for a population of $10^3$ BL Lac sources with a cut-off of $10^{14}\ \rm{eV}$. Dashed lines represent the same emission without cut-offs. Red dots are the current upper limits set by the Pierre Auger Observatory.}
    \label{fig:bllac:limits2}
\end{figure}

Using the framework developed in the main text, we applied the synchrotron and synchrotron self-Compton (SSC) models to the case of Mrk~501, adopting representative source parameters from the literature \cite{Albert_2021}. The resulting synchrotron spectra, shown in Figures~\ref{fig:mkr_sync} and~\ref{fig:mkr_sync_zoom}, indicate that LIV effects can extend the emission beyond the standard LI cutoff and even modify the low-energy part of the spectrum for sufficiently large LIV parameters. In the SSC channel, the predicted spectra displayed in Figure~\ref{fig:mkr} reveal the emergence of an additional high-energy emission region that is absent in LI models. Extending the same procedure to a simplified population of BL Lac objects, we modeled each source using typical parameters reported in Ref.\cite{Zhang_2013}, and then computed the cumulative photon spectrum as the simple sum of contributions from approximately $10^3$ known BL Lac objects \cite{FermiLAT_2019,Ballet_2023}. This cumulative photon spectrum reaches energies above $10^{17} \ \rm{eV}$, as illustrated in Figures~\ref{fig:bllac:spectrum1_sync}–\ref{fig:bllac:spectrum2_sync} for synchrotron emission and in Figures~\ref{fig:bllac:spectrum1}–\ref{fig:bllac:spectrum2} for SSC emission. For reference, these fluxes are also compared with the upper limits on the ultra-high-energy photon flux reported by the Pierre Auger Observatory in Figures~\ref{fig:bllac:limits1_sync}–\ref{fig:bllac:limits2_sync} (synchrotron) and Figures~\ref{fig:bllac:limits1}–\ref{fig:bllac:limits2} (SSC).
All fluxes shown were computed in the vicinity of the sources, so propagation effects were not taken into account.

We stress, however, that these results are only meant as an illustration. Deriving robust limits on LIV coefficients from this type of analysis would require a series of additional steps, including: (i) implementing a full dynamical treatment of synchrotron and inverse Compton processes, rather than a purely kinematical one; (ii) modeling photon propagation with LIV, including absorption in the extragalactic background light and possible modifications to pair production thresholds; (iii) accounting for the development of atmospheric showers at ultra-high energies, which directly impacts observational signatures; (iv) incorporating realistic source populations with distributions in distance, luminosity, and variability; and (v) considering other relevant source classes such as flat-spectrum radio quasars. Each of these ingredients can alter the resulting spectrum in a nontrivial way and must be included before quantitative bounds can be established.

Within these limitations, the present appendix should be regarded as a first step, illustrating the qualitative trends expected when LIV is incorporated into blazar emission models. A complete treatment combining dynamics, propagation, and population effects will be required in future work to transform these signatures into quantitative constraints on LIV parameters.


\end{document}